\documentclass[sigconf]{acmart}
\renewcommand{\footnotetextcopyrightpermission}[1]{}
\usepackage{graphicx}
\usepackage{tabularx}
\usepackage{booktabs}
\usepackage{multirow}
\usepackage{hyperref}
\usepackage[most]{tcolorbox}
\tcbuselibrary{listings, skins}
\tcbuselibrary{raster}
\usepackage{enumitem}
\usepackage{xcolor}

\definecolor{keywordcolor}{RGB}{0,0,180}
\definecolor{functioncolor}{RGB}{0,100,0}
\definecolor{stringcolor}{RGB}{180,0,180}

\setlist{topsep=2pt,itemsep=1pt,parsep=0pt}

\newtcolorbox{findingsbox}{
  breakable,
  colback=gray!10,
  colframe=gray!50,
  boxrule=0.5pt,
  arc=2pt,
  left=6pt,
  right=6pt,
  top=6pt,
  bottom=6pt
}

\lstdefinestyle{pythonstyle}{
    language=Python,
    basicstyle=\ttfamily\footnotesize,
    keywordstyle=\color{keywordcolor}\bfseries,
    stringstyle=\color{stringcolor},
    commentstyle=\color{gray},
    numbers=left,
    numberstyle=\tiny\color{gray},
    stepnumber=1,
    numbersep=8pt,
    framesep=4pt,
    backgroundcolor=\color{white},
    showstringspaces=false,
    breaklines=true,
    tabsize=4,
    captionpos=b,
    frame=lines,
    literate={torch.compile}{torch.\normalcolor compile}{13}
}

\usepackage[commandnameprefix=ifneeded]{changes}

\definechangesauthor[color=blue]{wei}

\definechangesauthor[color=red]{xinyi}

\definechangesauthor[color=purple]{jiaxin}

\definechangesauthor[color=orange]{jinyi}

\newcommand{\issue}[1]{\href{https://github.com/pytorch/pytorch/issues/#1}{\##1}}

\newcommand{\cissue}[1]{\colorbox{gray!20}{\textbf{\issue{#1}}}}

\title{Demystifying Deep Learning Compiler Frontend Bugs: An LLM-Aided Empirical Study}

\author{Xinyi Yuan}
\affiliation{%
  \institution{Institute of Software, Chinese Academy of Sciences}
  \institution{University of Chinese Academy of Sciences}
  \state{Beijing}
  \country{China}
}
\email{yuanxinyi22@otcaix.iscas.ac.cn}

\author{Wei Chen}
\authornote{Corresponding author.}
\affiliation{%
  \institution{Institute of Software, Chinese Academy of Sciences}
  \institution{University of Chinese Academy of Sciences}
  \state{Beijing}
  \country{China}
}
\email{wchen@otcaix.iscas.ac.cn}

\author{Jinyi Liu}
\affiliation{%
  \institution{Institute of Software, Chinese Academy of Sciences}
  \institution{University of Chinese Academy of Sciences}
  \state{Beijing}
  \country{China}
}
\email{liujinyi24@otcaix.iscas.ac.cn}

\author{Pengyu Chen}
\affiliation{%
  \institution{Institute of Software, Chinese Academy of Sciences}
  \institution{University of Chinese Academy of Sciences}
  \state{Beijing}
  \country{China}
}
\email{chenpengyu23@otcaix.iscas.ac.cn}

\author{Jun Wei}
\affiliation{%
  \institution{Institute of Software, Chinese Academy of Sciences}
  \institution{University of Chinese Academy of Sciences}
  \state{Beijing}
  \country{China}
}
\email{wj@otcaix.iscas.ac.cn}

\author{Guoquan Wu}
\affiliation{%
  \institution{Institute of Software, Chinese Academy of Sciences}
  \institution{University of Chinese Academy of Sciences}
  \state{Beijing}
  \country{China}
}
\email{gqwu@otcaix.iscas.ac.cn}

\author{Jiaxin Zhu}
\affiliation{%
  \institution{Institute of Software, Chinese Academy of Sciences}
  \institution{University of Chinese Academy of Sciences}
  \state{Beijing}
  \country{China}
}
\email{zhujiaxin@otcaix.iscas.ac.cn}

\author{Tao Huang}
\affiliation{%
  \institution{Institute of Software, Chinese Academy of Sciences}
  \institution{University of Chinese Academy of Sciences}
  \state{Beijing}
  \country{China}
}
\email{tao@otcaix.iscas.ac.cn}

\begin{document}

\begin{abstract}


Deep learning compilers (DLCs) are designed to translate deep learning programs into optimized, hardware-specific code. Typically, DLC frontends translate programs into graph-based intermediate representations (IRs) to enable optimizations. Defects introduced during this stage (termed \emph{fBug}s) are severe yet understudied, as prior work predominantly focuses on low-level APIs and operators or treats DLCs as monolithic entities. 

To bridge this gap, we conduct the first systematic empirical study of \emph{fBug}s in TorchDynamo, the default DLC frontend for PyTorch 2, the most popular DL framework. Leveraging a domain-knowledge-enhanced LLM-aided methodology, we analyze 123 \emph{fBug}s and construct a taxonomy comprising 7 root cause categories and 15 subcategories. Our findings provide actionable insights for DLC development and testing. Furthermore, we leverage the LLM to generate targeted, root cause-aware test cases to detect new bugs. We uncovered 23 previously unknown \emph{fBug}s in recent releases (15 confirmed) across eight (sub)categories, demonstrating the efficacy of our methodology in testing and hardening DLC frontends.

\end{abstract}


\keywords{Deep Learning Compiler; Frontend Bug; PyTorch; TorchDynamo; LLM; Empirical Study}

\maketitle

\section{Introduction}
The rapid adoption of deep learning (DL) has driven the development of deep learning compilers (DLCs), which bridge the gap between high-level frameworks (e.g., PyTorch~\cite{2019PyTorchPaper, 2024PyTorchPaper} and TensorFlow~\cite{2016TensorFlowPaper}) abstractions and hardware-specific optimizations. DLCs, such as TVM~\cite{TVM2018}, Glow~\cite{Glow2018}, and AKG~\cite{2021AKG}, typically consist of a frontend and a backend. The frontend captures the computational graph by translating high-level DL models into an intermediate representation (IR) and applies graph-level optimizations like operator fusion and constant folding; the backend further lowers the IR into hardware-specific operators, ensuring efficient execution. By decoupling algorithm design from hardware implementation, DLCs enhance both usability and performance, making them critical for the reliability and efficiency of DL frameworks. 

However, defects in DLCs can have severe consequences. They may propagate through the compilation pipeline and lead to crashes, incorrect outputs, or performance degradation. While prior studies have tested DLCs and investigated their bugs, most focus on backend components, such as operator implementations or hardware-specific optimizations. The frontend’s program-to-graph translation and corresponding errors remain relatively underexplored, despite being equally critical. To bridge this gap, we conduct the first empirical study on the bugs (i.e., \emph{\textbf{fBug}s}) of the compiler frontend \textbf{TorchDynamo} in the most popular DL framework, PyTorch 2~\cite{2024PyTorchPaper}. We aim to demystify the characteristics of \emph{fBug}s in terms of their distributions, triggers, and root causes. 

We propose an LLM-aided empirical study that addresses key challenges in analyzing complex, real-world bug reports. To guide the LLM in extracting critical information from noisy, unstructured bug reports and patches, we design a specialized template to characterize \emph{fBug} features. To mitigate LLM hallucination and improve reliability, we further inject domain knowledge of TorchDynamo as contextual grounding. Building upon the LLM's analysis, we manually review and consolidate the results to distill the root causes of bugs into a fine-grained classification system, which serves as a structured reference for downstream tasks. Leveraging information on bug root causes, an LLM can generate new test cases to detect bugs in the DLC frontend of the recent version.

Following the methodology, our in-depth investigation of 123 \emph{fBug}s identifies 7 root cause categories and 15 subcategories. We obtain several key insights: inadequate support in the DLC frontends for complex Python features, e.g., dynamic typing, meta-programming, and runtime introspection, is a major source of \emph{fBug}s; \emph{fBug}s sharing the same root cause tend to be triggered by similar Python constructs and runtime conditions; and identified root causes enable the LLM to generate test cases for new \emph{fBug}s. Our LLM-aided method detects 23 \emph{fBug}s in later DLC versions, with 15 confirmed across eight (sub)categories. 

The main contributions are summarized as follows:
\begin{itemize}[leftmargin=*]
    \item We conduct the first systematic empirical study of DLC \emph{fBug}s in the most popular DL framework PyTorch, yielding valuable insights into bug root causes and characteristics.
    \item We design a novel LLM-aided empirical study methodology, which can (1) analyze \emph{fBug}s accurately and summarize them reasonably based on the instructions enriched with DLC domain knowledge and (2) effectively create test cases, based on root causes, that can trigger previously unseen \emph{fBug}s.
    \item This work offers valuable takeaways for the development, use, and testing of the DLC frontend. Source code and datasets are publicly available for reproducibility and future research.
\end{itemize}

The rest of this paper is organized as follows.
Section~\ref{sec:DLC} introduces DLCs and PyTorch’s compilation pipeline. Section~\ref{sec:Motivation} presents our motivation and explains our focus on TorchDynamo. Section~\ref{sec:Methodology} outlines our research questions and proposes the LLM-aided methodology. Section~\ref{sec:StudyResult} details the study results, and Section~\ref{sec:Discussion} discusses the role of LLMs, implications, limitations, and threats to validity. Section~\ref{sec:RelatedWorks} reviews related work, and Section~\ref{sec:Conclusion} concludes the paper.

\section{Deep Learning Compiler}
\label{sec:DLC}
DLCs  bridge high-level programming frameworks and diverse hardware, translating DL models into optimized, hardware-specific code that improves parallelism and memory efficiency. 
A rich ecosystem of DLCs has emerged, including TVM~\cite{TVM2018}, Glow~\cite{Glow2018}, XLA~\cite{2017XLA}, nvFuser~\cite{nvFuser}, AKG~\cite{2021AKG}, and TorchDynamo with TorchInductor~\cite{2024PyTorchPaper}.

PyTorch 2 has emerged as a leading DL framework, widely adopted in both academic research and industry for its balance of usability and performance~\cite{PTPopular}. A key advancement in PyTorch 2 is its staged compilation pipeline, activated via \texttt{torch.compile}, which addresses the limitations of prior eager and graph execution modes. The pipeline consists of three core components — \textit{TorchDynamo}, \textit{AOTAutograd}, and \textit{TorchInductor} — each responsible for optimizing execution at a distinct stage. Specifically,

\textit{\textbf{TorchDynamo}} acts as the DLC frontend. It captures the computational graph by simulating Python bytecode execution and generating an \textbf{FX graph}~\cite{2022torchFXPaper} that encodes the forward computation and necessary runtime constraints.

\textit{\textbf{AOTAutograd}} takes the FX graph as input and applies ahead-of-time automatic differentiation to construct explicit backward computation graphs. Both forward and backward graphs serve as the input to backend compilation.

\textit{\textbf{TorchInductor}} is the default compiler backend that consumes the forward and backward graphs and lowers them to hardware-specific code. It applies backend-level optimizations such as operator fusion and memory planning, and generates executable kernels for CPUs and GPUs (e.g., via Triton~\cite{2019triton} or C++).

\begin{figure}[t]
    \centering
    \includegraphics[width=0.75\linewidth]{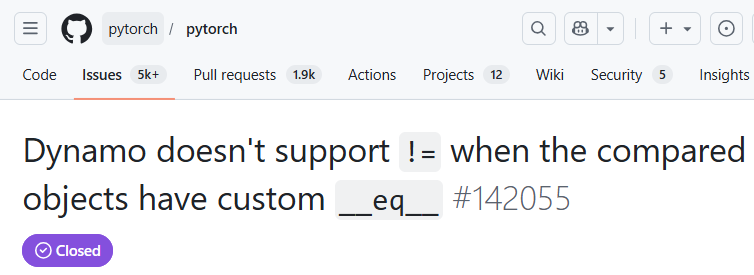}
    \caption{An Issue of PyTorch Related to DLC Frontend}
    \label{fig:pt142055}
\end{figure}

\section{Motivation and Object}
\label{sec:Motivation}
This section analyzes the problem and explains why we target TorchDynamo.

\subsection{Problem Analysis}
DLCs are crucial to the correctness and performance of DL programs, but some concerns still remain.
First, prior studies on DLC bugs often treat DLCs as monoliths, neglecting the distinct roles of the frontend and backend. This obscures stage-specific failure modes and hinders a deeper understanding of bugs.
Second, existing work focuses predominantly on backend components, leaving the frontend's program-to-graph translation  understudied. However, errors from the frontend are equally critical as they can propagate throughout the entire compilation pipeline.



Figure~\ref{fig:pt142055} illustrates a representative \emph{fBug}: a user-defined \texttt{\_\_eq\_\_} is silently ignored with \texttt{torch.compile}, causing \texttt{!=} to fall back to the default built-in behavior and producing incorrect outputs. This motivates a systematic empirical study of \emph{fBug}s, focusing on their root causes, triggers, and implications for DLC reliability.

\begin{figure*}[ht]
    \centering
    \includegraphics[width=0.95\linewidth]{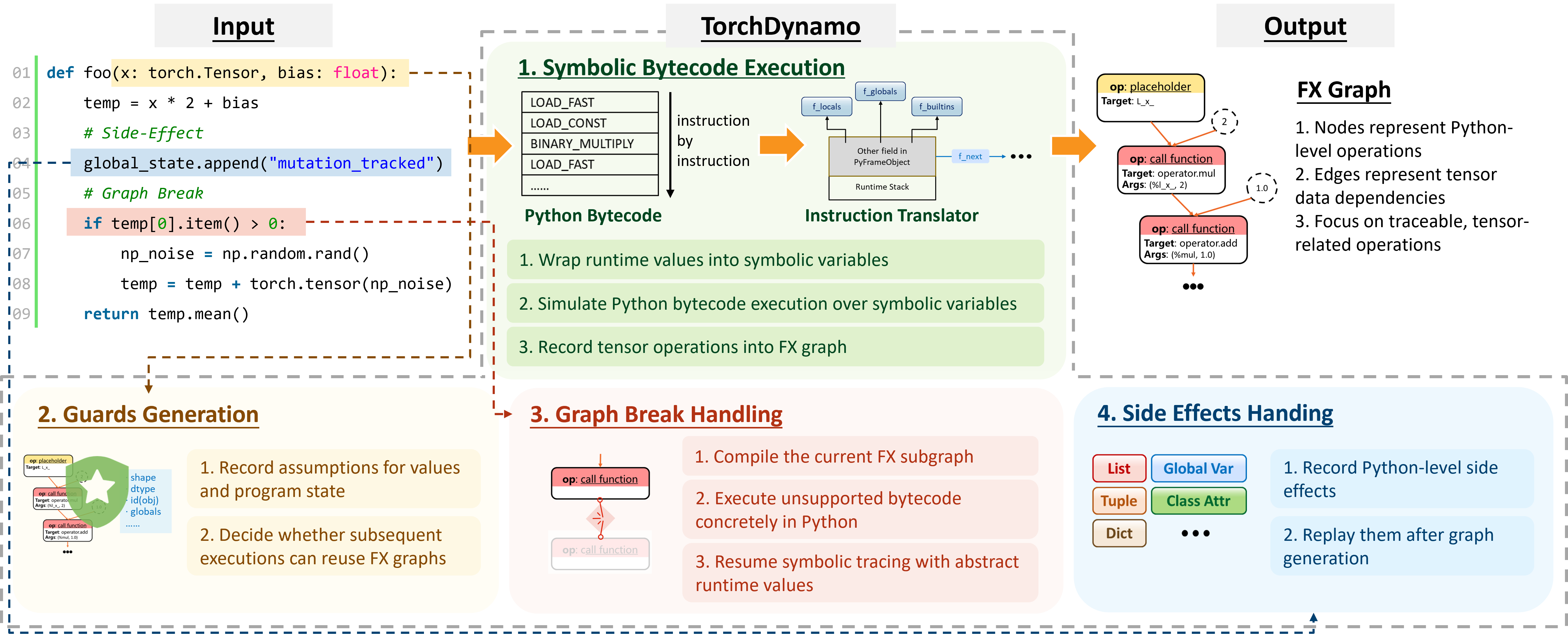}
    \caption{TorchDynamo Workflow}
    \label{fig:dynamo_overview}
\end{figure*}

\subsection{TorchDynamo}
\label{subsec:TorchDynamo}

We choose TorchDynamo as our research object for three key reasons. First, PyTorch's dominance in both research and industry makes the correctness of TorchDynamo, its default compiler frontend, critically important to the deep learning community. Second, TorchDynamo exemplifies a sophisticated Just-In-Time (JIT) compiler frontend that bridges the gap between dynamic, high-level languages (e.g., Python) and static performance optimization. Finally, its execution model introduces distinct and complex failure modes, and its rapid evolution and numerous issues provide ample research cases.

\section{Methodology}
\label{sec:Methodology}

We attempt to answer the following three research questions.
\begin{itemize}[leftmargin=*]
    \item \textbf{RQ1 (Distribution):} Are \emph{fBug}s prevalent? What are their characteristics in terms of distributions and symptoms?
    \item \textbf{RQ2 (Root cause):} What are the root causes of \textit{fBug}s?
    \item \textbf{RQ3 (Bug detection):} Are these root causes helpful in discovering new bugs?
\end{itemize}

\begin{figure*}[t]
    \centering
    \includegraphics[width=0.9\linewidth]{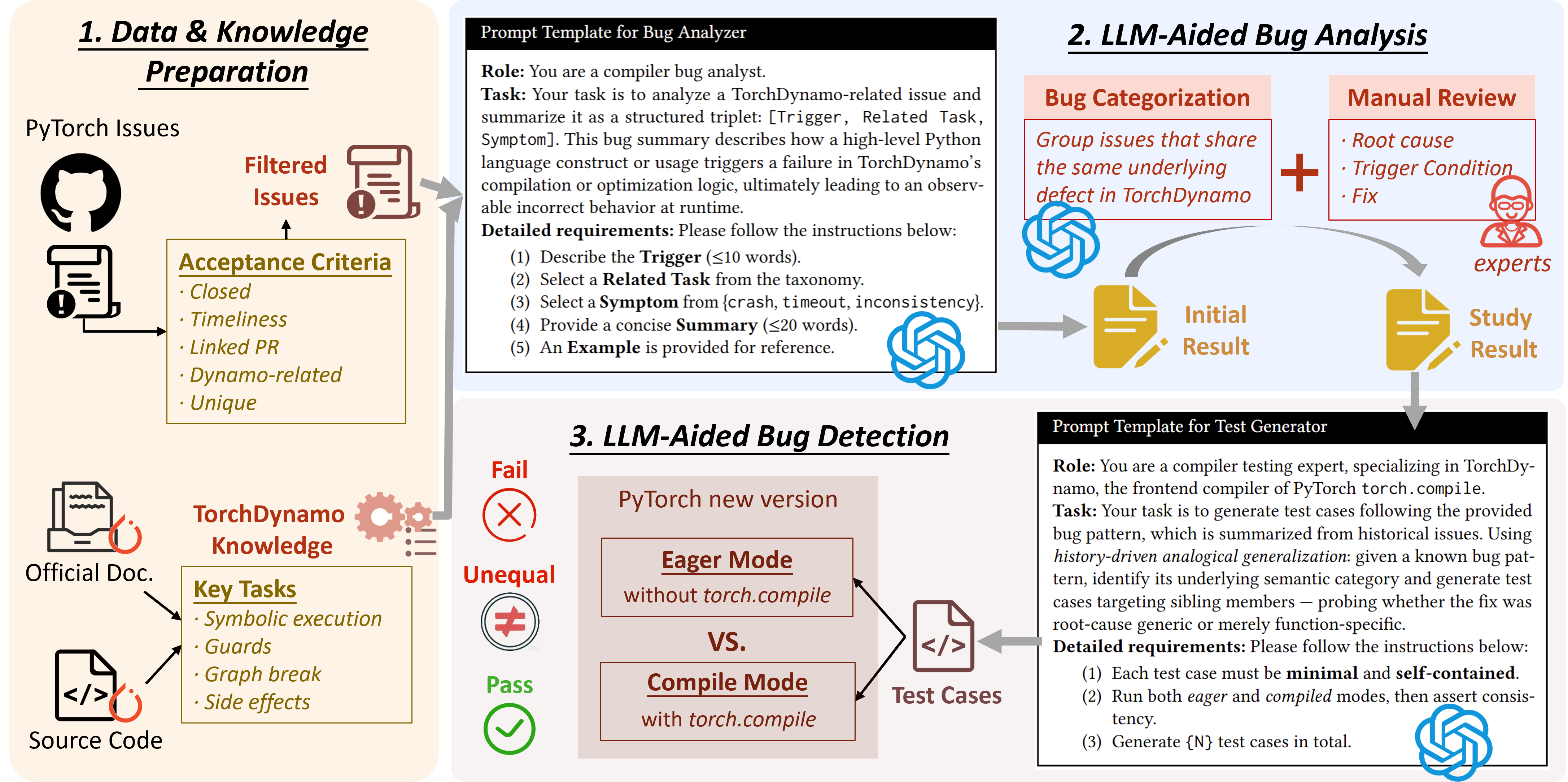}
    \caption{Methodology Overview of Our LLM-Aided Empirical Study}
    \label{fig:Workflow}
\end{figure*}

We design an LLM-aided empirical study methodology. 
As illustrated in Figure~\ref{fig:Workflow}, our study involves three stages: (1) collecting high-quality \emph{fBug}s from real-world issue reports and distill DLC frontend specific knowledge, (2) analyzing these bugs by leveraging LLMs' code understanding and summarization capabilities, and (3) generating new test cases with an LLM, guided by our identified root causes, to detect similar bugs in new version.

\subsection{Data \& Knowledge Preparation}
\label{subsec:data_collection}
\subsubsection{Data preparation}
We obtain high-quality experimental data using a two-phase collection process, starting with automatic issue filtering via the GitHub API. The filtering criteria are as follows.

\begin{itemize}[leftmargin=*]
    \item \textbf{Closed issues.} 
    We focus on closed issues, as their resolutions provide definitive outcomes for our analysis.

    \item \textbf{Timeliness.} The issues should be between \emph{2024-06-30} and \emph{2025-06-30}, ensuring that these bugs arose in modern PyTorch 2.x with TorchDynamo.

    \item \textbf{TorchDynamo relevance.} 
    We retain only issues labeled \textbf{\emph{module: dynamo}} that constitute reports related to \texttt{torch.compile}.

    \item \textbf{Code changes for bug fix.} 
    Each issue must link to at least one pull request (PR) that fixes the bug, enabling precise tracking of code changes and analysis of repair strategies. 
\end{itemize}

Further, we manually review all candidate issues to eliminate potential irrelevant cases or noise. We exclude non-defect cases (e.g., API misuse, expected behaviors, or defects from other components), duplicates identified through similar error signatures and developer discussions, as well as feature requests and performance discussions. This process yields a final dataset of \textbf{123} confirmed, unique, and relevant TorchDynamo bugs across PyTorch 2.1 to 2.6, providing a reliable basis for subsequent analysis.

\subsubsection{Knowledge preparation}
By analyzing the documentation and source code, we gain a comprehensive understanding of TorchDynamo, focusing on its workflow and tasks. We observe that TorchDynamo symbolically models DL program execution instruction by instruction to translate operations into an FX graph. Further, we abstract its workflow, illustrated in Figure~\ref{fig:dynamo_overview}, which consists of four key tasks.

(1) \textbf{Symbolic bytecode execution.}
This is a fundamental task in the entire workflow. TorchDynamo uses \emph{VariableTracker}s, symbolic proxies, to wrap runtime variables and record behaviors of various data types~\cite{2024PyTorchPaper}. It simulates Python bytecode execution using these proxies instead of concrete values, tracking the frame state, including the stack state and local variables. When encountering a PyTorch operation, TorchDynamo creates an FX node with edges representing data dependencies between operations, incrementally building a graph that represents program execution.

(2) \textbf{Guards generation.}
TorchDynamo uses \textit{guards} to verify that runtime assumptions made during symbolic execution, such as tensor metadata (e.g., shape, dtype), remain valid, enabling the reuse of cached compilations. These guards are checked before executing a compiled graph, and recompilation is triggered if any guard fails. For instance, as shown in Figure~\ref{fig:dynamo_overview}, TorchDynamo creates guards (Line 1) to verify that \texttt{x} has specific tensor metadata and that \texttt{bias} is a float, ensuring the cached graph's validity for the current inputs.

(3) \textbf{Graph break handling.}
Upon encountering a Python bytecode that cannot be handled, TorchDynamo breaks the graph, compiles the captured subgraph up to that point, and returns to the standard Python interpreter. For example, in Figure~\ref{fig:dynamo_overview}, the data-dependent control flow at Line 6 causes a graph break because \texttt{temp}'s concrete value cannot be known in compilation. This allows TorchDynamo to compile compatible regions incrementally while falling back to eager execution elsewhere.

(4) \textbf{Side effects handling.}
Side effects are observable updates to program state (globals, attributes, containers, or closures) that FX cannot directly represent. TorchDynamo records these updates during symbolic execution and defers applying them until FX graph execution, materializing the necessary ones in the output bytecode. For example, in Figure~\ref{fig:dynamo_overview}, Line 4 shows an external object mutation handled as a side effect.

Our analysis reveals that (1) \textit{various data structures}, \textit{guards}, \textit{FX graph}, and \textit{side effects} are critical entities in the compilation pipeline; (2) these four tasks are essential for compilation, but their complexities can introduce bugs.

We structure the analysis and summary into a well-defined format of named tasks with descriptions, creating DLC-frontend-specific knowledge that helps LLMs avoid ambiguity and knowledge conflicts.

\subsection{LLM-Aided Bug Analysis}
We propose an LLM-human collaboration approach for comprehensive \emph{fBug} analysis, decomposed into two steps: issue report analysis and root cause categorization. This approach leverages LLMs to efficiently understand and summarize bugs, complemented by human expertise for validation.

To extract structured information from unstructured issue reports, each issue is analyzed by an LLM (GPT-5~\cite{2025GPT5card}) and characterized by four fields:  \emph{related task}, \emph{trigger}, \emph{symptom}, and \emph{summary}, capturing \textit{when}, \textit{why}, and \textit{what} an issue manifests. This structured representation provides a uniform format suitable for subsequent statistical analysis and categorization.

The \emph{related task} is selected from the four tasks identified in Section~\ref{subsec:data_collection}, prompting the LLM to analyze each issue in the context of TorchDynamo's core functionalities. The \emph{symptom} is drawn from three enumerated categories---\emph{crash}, \emph{timeout}, and \emph{inconsistency}---consistent with the classification scheme adopted in prior empirical studies~\cite{2021SurveyShen, 2021SurveyDu,2025SurveyHuang}. 
The \emph{trigger} and \emph{summary} are expressed in natural language. To ensure consistent outputs and reduce variance, we restrict output length and provide a one-shot example.

With these 123 initial analysis results, we cluster bugs by root cause and further divide each cluster by trigger. 
To ensure reliable results, we query two LLMs (GPT-5~\cite{2025GPT5card} and Claude-Sonnet-4~\cite{2025Claude4card}) and involve two authors to review all LLM-generated annotations. All disagreements are resolved through manual inspection and discussion until consensus is reached.


\subsection{LLM-Aided Bug Detection}
\label{subsec:discovery}
We employ an LLM (GPT-5~\cite{2025GPT5card}) to generate new test cases that are semantically similar to known bugs but target different situations. Specifically, we design a root cause-aware test case generation strategy. For each identified root cause, we incorporate its name and a description explaining how such bugs occur with respect to the program structure, semantic features, related entities, and functions in an LLM prompt. In addition, we impose several constraints on the generated tests. Each test must be self-contained, as simple as possible, and executable with and without \texttt{torch.compile}. We also prompt the LLM to generate corresponding assertions to compare the execution results. Finally, we execute each generated test case in a recent version of PyTorch\footnote{We use PyTorch 2.10 in this experiment.}, following the idea of differential testing~\cite{1998DiffTest}, i.e., comparing results between eager and compile modes.  This allows us to detect potential regressions and assess whether semantically similar bugs can still be triggered.

\section{Study Result}
\label{sec:StudyResult}
This section presents our empirical study results and the answers to the research questions. Beyond that, we draw multiple valuable findings and implications.

\subsection{Bug Distribution (RQ1)}

\begin{table}[htb]
\centering
\small
\setlength{\tabcolsep}{4pt}
\caption{Distribution of \emph{fBug}s}
\begin{tabularx}{\linewidth}{lXXXXX}
\hline
 & \textbf{Symbolic Exec.} 
 & \textbf{Graph Break} 
 & \textbf{Side Effects} 
 & \textbf{Guards} 
 & \textbf{Total}
 \\
\hline
custom class           & 20 & 0  & 6  & 6 & 32 \\
container              & 14 & 1  & 0  & 1 & 16 \\
iterator               & 5  & 0  & 0  & 0 & 5  \\
closure                & 1  & 4  & 2  & 1 & 8  \\
Python scalar          & 4  & 1  & 0  & 3 & 8  \\
tensor op              & 4  & 0  & 1  & 3 & 8  \\
decorator              & 1  & 0  & 1  & 0 & 2  \\
global state           & 0  & 0  & 6  & 1 & 7  \\
in-place op            & 0  & 0  & 5  & 0 & 5  \\
ctx manager            & 0  & 11 & 0  & 1 & 12 \\
compile cfg            & 0  & 1  & 0  & 1 & 2  \\
unsup. op              & 8  & 3  & 0  & 0 & 11 \\
others                 & 0  & 3  & 2  & 2 & 7  \\
\hline
\textbf{Total}            & \textbf{57} & \textbf{24} & \textbf{23} & \textbf{19} & \textbf{123}\\
\hline
\end{tabularx}
\label{tab:dynamo_mechanism_vs_trigger}
\raggedright\noindent\footnotesize \emph{tensor op}, \emph{in-place op}, and \emph{unsup.~op} refer to tensor operations, in-place operations, and unsupported operations, respectively.
\end{table}

Table~\ref{tab:dynamo_mechanism_vs_trigger} summarizes the distribution of 123 \emph{fBug}s across the four key tasks and related entities, based on our understanding of TorchDynamo (Section~\ref{subsec:data_collection}). 

From the task perspective, \emph{symbolic execution} accounts for the majority of defects (57/123), significantly more than \emph{graph break} (24/123), \emph{side effects} (23/123), and \emph{guards} (19/123). This aligns with symbolic execution's complexity and foundational role: as the backbone of graph capture, it is both more error-prone and more frequently exercised, thus accounting for a large share of bugs.

From the entity perspective, \textit{custom classes} (32) and \textit{containers} (16) are the entities most susceptible to errors, primarily during \textit{symbolic execution}, with fewer issues in \textit{guards} and \textit{side effects}. This suggests that TorchDynamo graph capture struggles significantly with dynamic custom class objects (e.g., method resolution and introspection) and containers (e.g., lists and tuples).

The \emph{fBug} distribution reveals strong correlations between tasks and entities. For example, \emph{global state} (6/7) and \emph{in-place op} (5/5) issues are largely confined to \emph{side effects}, concerning the tracking and replaying of side effects. Iterator-related issues appear exclusively in symbolic execution (5/5), reflecting the difficulty of simulating this data structure’s execution. \emph{Closure} and \emph{Others} are exceptions, distributed across multiple tasks due to interaction bugs not attributable to a single mechanism.

In our analysis of the symptoms of the 123 bugs, \emph{crashes} dominate (88/123), indicating that most reported bugs manifest as immediate execution breakdowns. \emph{Inconsistencies} (28/123), though fewer, are equally critical because silent errors are difficult to detect while having serious consequences. \emph{Timeouts} are rare (7/123), and mostly (6/7) related to guards. Overall, reported TorchDynamo bugs tend to be explicit failures rather than silent errors. This is consistent with expectations, as the former are easier to detect and report.

\begin{findingsbox}
\textbf{Finding \#1}: 
\emph{fBug}s concentrate in \emph{symbolic execution} (57/123), particularly with high-level Python abstractions such as \textit{custom classes} and \textit{containers}, and exhibit strong correlations between tasks and related entities.
\end{findingsbox}

\begin{findingsbox}
\textbf{Implication \#1}: DLC frontends should strengthen the symbolic simulation of Python's custom objects. The entity-task correlations can serve as indicators to guide bug localization.
\end{findingsbox}

\begin{table*}
\centering
\small
\caption{Taxonomy of \emph{fBug} Root Causes}
\label{tab:bug-pattern}
\begin{tabularx}{\linewidth}{XXl}
\toprule
\textbf{Category} & \textbf{Subcategory} & \textbf{Num} \\
\midrule
A. Wrong modeling of Python objects
& A.1 Incorrect tracing of object initialization & 3\\
& A.2 Incomplete modeling of attribute access  & 8\\
& A.3 Overlooked magic method overrides & 7\\
& A.4 Misinterpretation of descriptors & 4\\
\midrule
B. Wrong modeling of containers
& B.1 Mis-modeled container abstraction \& instantiation & 9\\
& B.2 Untracked container mutation & 6\\
\midrule
C. Desynchronization of iterator state
& -- & 4\\
\midrule
D. Missing type transformation
& -- & 8\\
\midrule
E. Wrong handling of execution context \& scope
& E.1 Uncaptured context managers & 7\\
& E.2 Ignored exception handling & 3\\
& E.3 Wrong closure \& variable capturing & 5\\
& E.4 Untracked global state \& environment changes & 9\\
\midrule
F. Uncaptured side effects
& F.1 Untracked object attribute updates & 9\\
& F.2 Untracked tensor \& metadata updates & 7\\
\midrule
G. Guard overspecialization \& deficiencies
& G.1 Overspecialization of mutable or ephemeral state & 6\\
& G.2 Overspecialization of dynamic shapes & 5\\
& G.3 Missing or incomplete guards & 7\\
\midrule
H. Others & -- & 16 \\
\midrule
\textbf{Total} & -- & \textbf{123} \\
\bottomrule
\end{tabularx}
\end{table*}

\subsection{Root Cause (RQ2)}
\label{Root Causes}
Table~\ref{tab:bug-pattern} summarizes our taxonomy of root causes for the 123 \emph{fBug}s, comprising 7 categories and 15 subcategories.

\textit{\textbf{A. Wrong modeling of Python objects}}.
This category accounts for 18.7\% (23/123) of the bugs, making it a major root cause in our dataset. It is further divided into four subcategories.

\textit{A.1 Incorrect tracing of object initialization}.
TorchDynamo traces object operations symbolically, which can cause delegated or chained constructor calls to be skipped or only partially traced. This can lead to missing attribute definitions, potentially causing failures or inconsistent behavior during subsequent attribute accesses or mutations.
Listing~\ref{lst:A.1-Issue145284} exemplifies PyTorch Issue \#145284~\cite{PTIssue_145284}, where TorchDynamo invokes user-defined methods on \texttt{nn.Module} before it is fully initialized. This triggers an \texttt{AttributeError} because the symbolic tracer fails to track the internal fields \texttt{self.key\_cache}.

\begin{lstlisting}[style=pythonstyle, linewidth=0.9\linewidth, xleftmargin=0.05\linewidth, caption={Code Snippet of PyTorch Issue \#145284}, label={lst:A.1-Issue145284}]
class Cache(torch.nn.Module):
    def __init__(self):
        self.key_cache = []
    def __len__(self):
        return len(self.key_cache)
@torch.compile
def fn(x):
    cache = Cache()
    if cache:   # triggers __len__()
        return x
\end{lstlisting}

\textit{A.2 Incomplete modeling of attribute access}.
TorchDynamo parses attributes with fixed symbolic protocols, which may miss edge cases such as dynamic lookups and shadowed callables, resulting in incorrect internal state updates.
In Issue \#134844~\cite{PTIssue_134844} (Listing~\ref{lst:A.2-Issue134844}), the function \texttt{add\_one} is assigned as an attribute to the tensor \texttt{t}. However, this is not tracked by TorchDynamo and subsequent invocation of the callable attribute (i.e., \texttt{t.add\_one(t)}) fails.

\begin{lstlisting}[style=pythonstyle, linewidth=1.0\linewidth, xleftmargin=0.05\linewidth, caption={Code Snippet of PyTorch Issue  \#134844}, label={lst:A.2-Issue134844}]
@torch.compile
def foo(t):
    def add_one(x): ...
    t.add_one = add_one
    return t.add_one(t)
\end{lstlisting}

\textit{A.3 Overlooked magic method overrides}.
Python's support for magic methods and operator overloading allows users to customize object behavior, but TorchDynamo may overlook user-defined overrides, assuming default built-in behaviors.
In PyTorch Issue \#150265~\cite{PTIssue_150265} (Listing~\ref{lst:A.3-Issue150265}), \texttt{3 * x} triggers \texttt{\_\_rmul\_\_} on a custom tensor subclass. Due to the wrong logic of \texttt{has\_torch\_function}, TorchDynamo misclassifies the multiplication as a generic Python magic-method call rather than a tensor operator, causing an unexpected graph break.

\begin{lstlisting}[style=pythonstyle, linewidth=1.0\linewidth, xleftmargin=0.05\linewidth, caption={Code Snippet of PyTorch Issue \#150265}, label={lst:A.3-Issue150265}]
class MyParam(torch.nn.Parameter):
    def __new__(cls, data):
        return torch.Tensor._make_subclass(...)
@torch.compile
def f(x):
    return 3 * x   # triggers __rmul__
\end{lstlisting}

\textit{A.4 Misinterpretation of descriptors}. 
Python descriptors enable dynamic attribute binding by intercepting attribute access at runtime. When TorchDynamo fails to model this protocol correctly, it may simplify descriptor-backed attributes into plain functions or unbound methods based solely on static bytecode patterns. 
In PyTorch Issue \#155841~\cite{PTIssue_155841} (Listing~\ref{lst:A.4-Issue155841}),  the method \texttt{incr} is decorated with \texttt{functools.lru\_cache}. However, TorchDynamo models \texttt{f.incr} as a plain function rather than a bound method, thereby failing to implicitly bind \texttt{self} to the instance \texttt{f}. As a result, the call \texttt{f.incr(3)} lacks the instance argument, leading to an argument mismatch and compilation failure.

\begin{lstlisting}[style=pythonstyle, linewidth=1.0\linewidth, xleftmargin=0.05\linewidth, caption={Code Snippet of PyTorch Issue \#155841}, label={lst:A.4-Issue155841}]
class Foo:
    @functools.lru_cache
    def incr(self, val): ...
@torch.compile
def fn():
    Foo().incr(3)   # descriptor binding lost
    
\end{lstlisting}

\begin{findingsbox}
\textbf{Finding \#2}: 
A key reason for \emph{fBug}s is the gap between TorchDynamo’s symbolic execution and Python's dynamic object model, particularly in \emph{object initialization}, \emph{state mutation}, \emph{dynamic dispatch}, and \emph{runtime binding}.
\end{findingsbox}

\begin{findingsbox}
\textbf{Implication \#2}: DLC frontends must ensure full symbolic materialization of objects before they enter the tracing pipeline.
\end{findingsbox}

\textit{\textbf{B. Wrong modeling of containers}}.
This category comprises 15 bugs (12.2\%), rooted in a fundamental conflict in TorchDynamo: containers are not fully traced, yet their constituent tensors are first-class citizens. This triggers inconsistencies during container-dependent tensor access or control flow. We further divide this category into two subcategories based on container operation types.

\textit{B.1 Mis-modeled container abstraction \& instantiation}. 
This subcategory covers issues arising from container construction and abstraction, often triggered by automatic entry initialization, reflective access, and dynamic attribute binding (e.g., attaching functions or fields at runtime). 
In Issue \#141118~\cite{PTIssue_141118} (Listing~\ref{lst:B.1-Issue141118}), TorchDynamo misidentifies \texttt{MyWeirdDict}, a class with multiple inheritance, as a pure \texttt{nn.Module}. Thus, when accessed as a mapping, the call is erroneously routed through unsupported module-handling logic.

\begin{lstlisting}[style=pythonstyle, linewidth=1.0\linewidth, xleftmargin=0.05\linewidth, caption={Code Snippet of PyTorch Issue \#141118}, label={lst:B.1-Issue141118}]
class MyWeirdDict(collections.abc.MutableMapping, nn.Module):
    def __getitem__(self, item): ...
    def __setitem__(self, key, value): ...
    def __delitem__(self, item): ...
    def __len__(self): ...
    def __iter__(self): ...
@torch.compile
def to_weird_dict(td):
    return MyWeirdDict(**td)
\end{lstlisting}

\textit{B.2 Untracked container mutation}.
This subcategory captures issues caused by container mutations during execution. Representative cases include in-place updates and mutations under dynamic control flow (e.g., data-dependent iteration or exception paths). 
In Issue \#133063~\cite{PTIssue_133063} (Listing 6), a generator iterates over \texttt{self.mods} to extend list \texttt{y} while referencing \texttt{y[-1]}. Each iteration is expected to observe the latest state of \texttt{y}, but TorchDynamo fails to model this dynamic evolution and reads stale values.

\begin{lstlisting}[style=pythonstyle, linewidth=1.0\linewidth, xleftmargin=0.05\linewidth, caption={Code Snippet of PyTorch Issue \#133063}, label={lst:B.2-Issue133063}]
@torch.compile
class TempOpModel(nn.Module):
    def __init__(self):
        self.m = nn.ModuleList([...for _ in range(2)])
    def forward(self, x):
        y = [x, x]
        y.extend(m(y[-1]) for m in self.m)
        return y
\end{lstlisting}

\begin{findingsbox}
\textbf{Finding \#3}: 
TorchDynamo's inconsistent tracking of containers and their tensors causes container-related bugs, particularly when containers are dynamically modified, contain non-tensor elements, or are nested.
\end{findingsbox}

\begin{findingsbox}
\textbf{Implication \#3}: DLC frontends should adopt a unified, structure-aware tracking strategy that propagates monitoring through container boundaries.
\end{findingsbox}

\textit{\textbf{C. Desynchronization of iterator state}}.
This category, containing 4 bugs (3.25\%), arises from TorchDynamo’s inadequate modeling of iterator states during symbolic execution. Python iterators are highly flexible and stateful, making their behavior difficult to model symbolically. In Issue \#128944~\cite{PTIssue_128944} (Listing 7), for instance, a \texttt{for}-loop consumes a lazy \texttt{filter} iterator whose elements are consumed on demand. TorchDynamo fails to accurately capture the internal state transitions of such iterables, causing desynchronization between tracing and execution and triggering unexpected graph breaks.

\begin{lstlisting}[style=pythonstyle, linewidth=1.0\linewidth, xleftmargin=0.05\linewidth, caption={Code Snippet of PyTorch Issue \#128944}, label={lst:3-Issue128944}]
@torch.compile
def fn(inputs):
    it = filter(lambda t: t.requires_grad, inputs)
    out = inputs[0]
    for x in it:   # stateful consumption
        out = out * x
    return out
\end{lstlisting}

\begin{findingsbox}
\textbf{Finding \#4}: 
TorchDynamo's imprecise iterator modeling causes desynchronization between symbolic and concrete execution, especially for lazy iterables and generators with runtime-dependent evaluation.
\end{findingsbox}

\begin{findingsbox}
\textbf{Implication \#4}: DLC frontends should explicitly model iterator state, including materializing iterator outputs into trackable symbolic values, modeling iteration semantics, and maintaining state updates across loop boundaries.
\end{findingsbox}

\begin{lstlisting}[style=pythonstyle, linewidth=1.0\linewidth, xleftmargin=0.05\linewidth, caption={Code Snippet of PyTorch Issue \#156720}, label={lst:4-Issue156720}]
@torch.compile
class TestModel(torch.nn.Module):
    def forward(self, x, shape_params):
        return torch.ops.aten.view.default(x, shape_params)
x = torch.randn(24)
shape_params = [ torch.tensor(2, dtype=torch.int32),
                 torch.tensor(3, dtype=torch.int32),
                 torch.tensor(4, dtype=torch.int32) ]
\end{lstlisting}

\textit{\textbf{D. Missing type conversion}}.
This category contains 8 bugs (6.50\%), primarily due to inconsistent handling of tensors and Python primitives (e.g., scalar, \texttt{None}, boolean, string). The lack of explicit type normalization results in incorrect comparisons, arithmetic operations, and control-flow predicates. In Issue \#156720~\cite{PTIssue_156720} (Listing~\ref{lst:4-Issue156720}), scalars passed as shape parameters within a Python container cause an error because \texttt{aten.view.default} expects shape arguments of \texttt{SymInt} or \texttt{int}; however, the arguments remain unconverted \texttt{FakeTensor} objects, triggering an error.

\begin{findingsbox}
\textbf{Finding \#5}: 
TorchDynamo accurately proxies tensors but neglects crucial type and value conversions for non-tensors, leading to type errors in returns and comparisons.
\end{findingsbox}

\begin{findingsbox}
\textbf{Implication \#5}: DLC frontends should consider the types of non-tensor values, not just symbolic tensor proxies.
\end{findingsbox}

\textit{\textbf{E. Wrong handling of execution context \& scope}}.
This category contains 24 bugs out of 123 (19.5\%), the largest share in our dataset. We further divide it into four subcategories.

\begin{lstlisting}[style=pythonstyle, linewidth=1.0\linewidth, xleftmargin=0.05\linewidth, caption={Code Snippet of PyTorch Issue \#130559}, label={lst:E.1-Issue130559}]
@contextlib.contextmanager
def g(x):
    try:
        yield x.sin()
    finally:
        pass
@torch.compile
def fn(x):
    with g(x) as y:
        z = y + 1
    return z
\end{lstlisting}

\textit{E.1 Uncaptured context managers}.
In Python, context managers define runtime execution scopes via the \texttt{with} statement. Entering and exiting the \texttt{with} block is mandated by Python’s context management protocol. 
However, TorchDynamo may skip unrecognized context managers, incompletely emulate the \texttt{with}-block stack, or lose context state across graph breaks.
In Issue \#130559~\ref{lst:E.1-Issue130559} (Listing~\ref{lst:E.1-Issue130559}), the context manager is defined via \texttt{@contextlib.contextmanager}, which transforms a generator function into a context management object. However, this indirect definition of a context manager is not properly recognized and triggers an unexpected graph break.

\textit{E.2 Ignored exception handling}. 
Python uses \texttt{try/except} blocks to handle runtime errors and provide fallback logic. This subcategory covers issues in which TorchDynamo bypasses or mishandles the exception handler, skipping intended fallback paths.
For example, PyTorch Issue \#153605~\cite{PTIssue_153605} (Listing~\ref{lst:E.2-Issue153605}) is triggered by accessing a nonexistent attribute, which raises an \texttt{AttributeError}. Although this access occurs inside a \texttt{try/except} block, compilation crashes immediately instead of entering the \texttt{except} branch.

\begin{lstlisting}[style=pythonstyle, linewidth=1.0\linewidth, xleftmargin=0.05\linewidth, caption={Code Snippet of PyTorch Issue \#153605}, label={lst:E.2-Issue153605}]
@torch.compile
class Model(nn.Module):
    def forward(self, x):
        try:
            return torch.nn.functional.ATTR
        except AttributeError:
            return x
\end{lstlisting}

\textit{E.3 Wrong closure \& variable capturing}.
Python functions can access variables from enclosing scopes, such as local variables and nested functions. However, incorrect construction of closure environments during function inlining may lead to errors, particularly when closures interact with containers or global variables.
In Issue \#136814~\cite{PTIssue_136814} (Listing~\ref{lst:E.3-Issue136814}), nested function \texttt{fn} returns a closure \texttt{inner} that captures variable \texttt{x}. TorchDynamo is expected to inline both \texttt{fn} and \texttt{inner}, but incorrectly reconstructs the closure environment, triggering an unexpected \texttt{AttributeError} when accessing the captured variable.

\begin{lstlisting}[style=pythonstyle, linewidth=1.0\linewidth, xleftmargin=0.05\linewidth, caption={Code Snippet of PyTorch Issue \#136814}, label={lst:E.3-Issue136814}]
def test():
    x = torch.ones(1)
    def fn():
        def inner():
            return x + 2
        return inner
    @torch.compile
    def start():
        fn_inner = fn()
        res = fn_inner()
        return res, fn_inner
    start()
\end{lstlisting}

\textit{E.4 Untracked global state \& environment changes}.
Global variables and runtime configurations are frequently read and modified during program execution. However, to reduce the complexity of tracing, TorchDynamo’s symbolic evaluation initially assumes the global state is stable. This tracking strategy can fail to observe or propagate global state mutations, mishandle environment-dependent settings, or improperly apply cached graphs. 
In Issue \#132165~\cite{PTIssue_132165} (Listing~\ref{lst:E.4-Issue132165}), both functions in file \texttt{test\_import.py} modify the global variable \texttt{global\_flag}. In \texttt{repro.py}, the compiled function \texttt{fn} invokes them sequentially, mutating the same global variable twice. Since both calls originate from the same module object, TorchDynamo redundantly registers it in its side effect tracking system, leading to an assertion failure.

\begin{lstlisting}[style=pythonstyle, linewidth=1.0\linewidth, xleftmargin=0.05\linewidth, caption={Code Snippet of PyTorch Issue \#132165}, label={lst:E.4-Issue132165}]
# This repro requires two files
# file 1: test_import.py
global_flag = False
def set_flag_true():
    global global_flag
    global_flag = True
def set_flag_false():
    global global_flag
    global_flag = False

# file 2: repro.py
import test_import
@torch.compile()
def fn(x):
    test_import.set_flag_true()
    test_import.set_flag_false()
    return x + 1
\end{lstlisting}

\begin{findingsbox}
\textbf{Finding \#6}: 
Mishandling of execution context is the largest bug category, indicating TorchDynamo's tensor-centric design overlooks context-dependent state changes.
\end{findingsbox}

\begin{findingsbox}
\textbf{Implication \#6}: DLC frontends should precisely capture and restore execution contexts to avoid semantic divergence from Python runtime behavior.
\end{findingsbox}

\textit{\textbf{F. Uncaptured side effects}}.
This category contains 15 bugs (12.2\%) and is further divided into two subcategories.

\textit{F.1 Untracked object attribute updates}.
TorchDynamo may not correctly capture object attribute updates via non-standard patterns, e.g., aliased references or dynamic addition/deletion, leading to potential failures in deferred attribute writes.
In Issue \#143756~\cite{PTIssue_143756} (Listing~\ref{lst:F.1-Issue143756}), the constructor of \texttt{Something} mutates its attribute storage via a direct write to \texttt{self.\_\_dict\_\_}. This direct update to low-level mapping escapes TorchDynamo's side effect tracking, leading to an unexpected graph break during compilation.

\begin{lstlisting}[style=pythonstyle, linewidth=1.0\linewidth, xleftmargin=0.05\linewidth, caption={Code Snippet of PyTorch Issue \#143756}, label={lst:F.1-Issue143756}]
class Something:
    def __init__(self) -> None:
        self.__dict__["something"] = 'whatever'
@torch.compile
class MyModule(torch.nn.Module):
    def forward(self, x) -> torch.Tensor:
        Something()
        return x
\end{lstlisting}

\textit{F.2 Untracked tensor \& metadata updates}.
Stateful tensor updates may affect tensor metadata such as shape and dtype, particularly through in-place operations. TorchDynamo does not always anticipate these changes, resulting in incomplete modeling.
In Issue \#134820~\cite{PTIssue_134820} (Listing~\ref{lst:F.2-Issue134820}), a custom autograd function performs an in-place update via an \texttt{out=} argument. This write is expected to bypass autograd checks, but TorchDynamo inlines the \texttt{forward} method without properly modeling the no-grad boundary. Consequently, the \texttt{out}-variant observes an incorrect autograd state and raises an error.

\begin{lstlisting}[style=pythonstyle, linewidth=1.0\linewidth, xleftmargin=0.05\linewidth, caption={Code Snippet of PyTorch Issue \#134820}, label={lst:F.2-Issue134820}]
class toy_fn(torch.autograd.Function):
    @staticmethod
    def forward(ctx, x):
        torch.exp(x, out=x)
        return x
@torch.compile
def f(x):
    return toy_fn.apply(x)
\end{lstlisting}

\begin{findingsbox}
\textbf{Finding \#7}: Deferred side effect handling introduces semantic gaps. Inaccurate tracking or collapsing of updates by TorchDynamo can break ordering and state consistency with the original Python program.
\end{findingsbox}

\begin{findingsbox}
\textbf{Implication \#7}: DLC frontends should enforce precise ordering guarantees during side effect replay to preserve the original execution semantics.
\end{findingsbox}

\textit{\textbf{G. Guard overspecialization \& deficiencies}}.
This category contains 18 bugs (14.6\%), further divided into three subcategories.

\textit{G.1 Overspecialization of mutable or ephemeral state}.
TorchDynamo might over-specialize short-lived, instance-specific program states, misinterpreting them as stable constants.
In PyTorch Issue \#128319~\cite{PTIssue_128319} (Listing~\ref{lst:G.1-Issue128319}), \texttt{self.counter+=1} mutates a module attribute during each call. TorchDynamo inserts value-equality guards on this attribute, causing the guards to frequently invalidate and trigger repeated recompilation, which significantly increases compilation overhead.

\begin{lstlisting}[style=pythonstyle, linewidth=1.0\linewidth, xleftmargin=0.05\linewidth, caption={Code Snippet of PyTorch Issue  \#128319}, label={lst:G.1-Issue128319}]
class Mod(torch.nn.Module):
    def __init__(self):
        self.counter = 0
    def forward(self, x):
        self.counter += 1
        return self.linear(x)
opt_mod = torch.compile(Mod())
for _ in range(10):
    opt_mod(torch.randn(1, 1))
\end{lstlisting}

\textit{G.2 Overspecialization of dynamic shapes}. 
This subcategory covers bugs caused by overspecialization of tensor shapes. TorchDynamo may emit value-equality guards for dimensions that vary across inputs (e.g., batch sizes or mapped axes), effectively treating dynamic dimensions as constants. 
In Issue \#150540~\cite{PTIssue_150540} (Listing~\ref{lst:G.2-Issue150540}), a scalar tensor is converted to a concrete integer via \texttt{x.item()}, and shape constraints are checked using \texttt{\_check\_is\_size} and inequality guards. However, the symbolic relation between the derived index \texttt{b} and the constraint \texttt{b<y.shape[0]} is not preserved during export or compilation, leading to guard validation failures.

\begin{lstlisting}[style=pythonstyle, linewidth=1.0\linewidth, xleftmargin=0.05\linewidth, caption={Code Snippet of PyTorch Issue \#150540}, label={lst:G.2-Issue150540}]
@torch.compile
class M(torch.nn.Module):
    def forward(self, x, y):
        b = x.item()
        torch._check_is_size(b)
        torch._check(b < y.shape[0])
        return y[:b]
\end{lstlisting}

\textit{G.3 Absent or insufficient guards}. 
This allows compiled graphs to execute under invalid assumptions, leading to incorrect computations or unexpected exceptions.
In PyTorch Issue \#132692~\cite{PTIssue_132165} (Listing~\ref{lst:G.3-Issue132692}), the nested closure \texttt{forward\_function} captures the module field \texttt{self.linear}. During compilation, guard generation depends on a guard manager initialization path. When the flag \texttt{enable\_cpp\_guard\_manager} is disabled, the alternative guard path is taken without proper manager setup, leading to assertion failures during guard creation for closure-bound module variables.

\begin{lstlisting}[style=pythonstyle, linewidth=1.0\linewidth, xleftmargin=0.05\linewidth, caption={Code Snippet of PyTorch Issue \#132692}, label={lst:G.3-Issue132692}]
class LinearModel(torch.nn.Module):
    def forward(self, x):
        def forward_function(x):
            return self.linear(x)
        return forward_function(x)
torch._dynamo.config.enable_cpp_guard_manager = False
torch.compile(M())(...)
\end{lstlisting}

\begin{findingsbox}
\textbf{Finding \#8}: Inappropriate guard designs commonly cause \emph{fBug}s. Overspecialized or missing guards break the balance between correctness and reuse, leading to either excessive recompilation or incorrect graph reuse.
\end{findingsbox}

\begin{findingsbox}
\textbf{Implication \#8}: Guard mechanisms should precisely capture necessary assumptions while avoiding unnecessary specialization that harms reuse and performance.
\end{findingsbox}

\textit{\textbf{H. Others}}.
Notably, 16 issues are highly specific and too varied to form meaningful groups. For example, in PyTorch Issue \#144461~\cite{PTIssue_144461}, a PyTorch built-in benchmarking utility (i.e., \emph{ThroughputBenchmark}) unexpectedly mutates the global autocast dtype as a side effect of its internal execution, changing it from FP32 to BF16. TorchDynamo's global state guard then detects this mismatch and raises a \texttt{RecompileError}, aborting execution. The failure is triggered by an unexpected global state mutation introduced by an external benchmarking utility. Since no other similar APIs exhibit this behavior, it represents an isolated and specific bug.



\begin{table}[t]
\centering
\small
\caption{\emph{fBug}s Detected in Recent Release}
\label{tab:reproduce}
\begin{tabularx}{0.95\linewidth}{lX}
\toprule
\textbf{(Sub)category} & \textbf{New Issue} \\
\midrule
A.1 & \cissue{176596}, \issue{176692}\\
A.3 & \cissue{150765}, \cissue{174050}, \cissue{175841}, \cissue{176679}, \cissue{176686}, \issue{175615}, \issue{175943} \\
A.4 & \cissue{176599}\\
\midrule
C   & \cissue{175610}, \issue{175604}, \issue{175855}, \issue{176693} \\
\midrule
E.1 & \issue{176156} \\
E.2 & \cissue{174166}, \cissue{175608}, \cissue{175846}\\
E.4 & \cissue{175857}, \cissue{175953}, \issue{176252}\\
\midrule
F.1 & \cissue{176851}\\
F.2 & \cissue{176854}\\
\midrule
\textbf{Total} & \textbf{23}\\
\bottomrule
\end{tabularx}
\raggedright\noindent\footnotesize The 15 issues in \colorbox{gray!20}{\textbf{gray boxes}} are confirmed, the other 8 are pending.
\end{table}

\subsection{Bug Detection (RQ3)}
We observe that many bug fixes are case-specific, addressing only the immediate trigger rather than the complete underlying root cause. 
Take the Issue \#142055~\cite{PTIssue_142055} in Figure~\ref{fig:pt142055} as an example, PR \#142078~\cite{PTPR_142078} (\emph{[dynamo] Properly handle != under user-defined \_\_eq\_\_}) only addresses this particular case where \texttt{\_\_eq\_\_} or \texttt{\_\_ne\_\_} are overridden. This fix is narrowly scoped and motivates our LLM-aided detection for similar bugs beyond the exact patched condition.

We follow the method in Section~\ref{subsec:discovery} to generate ten test cases for each (sub)category in Section~\ref{Root Causes}.
We run all 170 test cases on PyTorch 2.10 and manually analyze each failure or inconsistency encountered.     
Overall, 23 out of the 170 test cases trigger failures or errors. We have reported these bugs, and 15 have been confirmed, covering 8 (sub)categories. Ten confirmed issues have been resolved with corresponding PRs. Table~\ref{tab:reproduce} lists our newly detected bugs.

Notably, five issues are tagged as \textbf{\emph{high priority}}, indicating that they have a significant impact on DLC stability and deserve urgent attention. Issue \#174166 was labeled as \textbf{\emph{good first issue}}, indicating that it provides insight into a new class of bugs. Issues \#174050 and \#176679 were tagged as \textbf{\emph{module: correctness (silent)}}, which is particularly concerning, because these bugs may produce incorrect results without raising any exceptions or warnings. 

Furthermore, the newly uncovered \textit{fBug}s reveal a prevalence of case-specific fixes over systematic root cause resolution. For instance, PR \#175611 only patched a specific \texttt{AttributeError} for \#174166, leaving the other two issues (\#175608, \#175846) with the same root cause (\textbf{E.2}) unresolved. This reactive patching, also evident in root cause \textbf{A.3}, suggests that systematic remediation remains challenging, especially when root causes stem from Python's dynamic language features.


We attribute the failure to discover new bugs in the remaining categories to four factors: (1) some categories require the interaction of multiple source files; (2) some others depend on specific external API calls from third-party libraries; (3) some are hardware-specific and only manifest in certain GPU environments; (4) and some involve APIs that have since been removed or renamed. These cases fall outside the scope of our current approach or cannot be reproduced in our environment.

\begin{findingsbox}
\textbf{Finding \#9}: 
TorchDynamo maintenance often relies on fragmented, case-specific patches. Nevertheless, the root-cause-aware test case generation is underscored by the discovery of new \emph{fBug}s in 47\% (8/17) of the subcategories.
\end{findingsbox}

\begin{findingsbox}
\textbf{Implication \#9}: 
Bug root causes can effectively guide an LLM to synthesize test cases in detecting new \emph{fBug}s. 
\end{findingsbox}

\section{Discussion}
\label{sec:Discussion}
\subsection{LLM Aids Empirical Study}
LLMs significantly accelerate our study, as evidenced by PyTorch Issue \#142055~\cite{PTIssue_142055} in Figure~\ref{fig:pt142055}.  While manual analysis, reading reports, running code, and reviewing fixes take approximately one hour, our LLM-based pipeline generates structured summaries in <30 seconds. Combined with <10 minutes of human verification, this approach yields an 80\% reduction in analysis time. Beyond efficiency, LLMs improve categorization consistency; guided by domain-knowledge augmented prompts, they reliably map complex issues (e.g., those involving Python magic methods) to potential root causes that human analysts might overlook.

Notably, our methodology is LLM-agnostic, as domain knowledge is explicitly encoded in the prompt. All LLM outputs undergo cross-validation and independent review by two authors, ensuring that the LLM serves as an accelerator rather than a decision-maker and that conclusions remain robust across different models.

\subsection{Implications}
\textbf{For DLC developers}: Our findings highlight the impedance mismatch between Python’s dynamic semantics and static graph requirements. To prevent silent invalidation of tracing, DLC frontends must move beyond ad hoc bug handling toward exhaustive state tracking and robust, semantic-aware guard mechanisms.

\textbf{For framework users}: To ensure reliable acceleration, users should adopt compiler-friendly programming, prioritizing functional purity and explicit state management to mitigate unexpected behavioral divergences during compilation.

\textbf{For testers}: Integrating domain knowledge, such as design assumptions and known bug root causes, with LLMs enables a scalable paradigm for synthesizing effective tests.

\subsection{Threats to Validity}
\textbf{Internal threat.}
To mitigate potential LLM hallucinations and ensure analysis accuracy, we integrate domain knowledge into prompt engineering and manually verify all LLM-generated outputs.

\textbf{External threat.}
While our study focuses on a single DLC, the identified root causes characterize graph IR transformations common across mainstream deep learning compilers. Furthermore, our LLM-aided methodology is transferable to other DLC frontends via prompt tuning.

\subsection{Limitations and Future Work}
First, our test case synthesis is guided by root causes from known \emph{fBug}s; for unseen bugs that fall outside the identified categories, the effectiveness of root-cause-aware generation is a concern. Second, as the DLC ecosystem evolves and more issues are reported, the stability and representativeness of our categorization may shift. Future work would extend the study using larger, more diverse datasets to further broaden the generalizability of our findings.

\section{Related Works}
\label{sec:RelatedWorks}
The most related work includes testing and empirical studies of DL systems, especially DL frameworks and compilers. We also briefly review LLM-aided empirical study and traditional compiler testing.

\textbf{Empirical studies on DL system bugs}.
Research in this domain bifurcates into DL frameworks and compilers. For DL frameworks, existing studies~\cite{2019SurveyIslam, 2020SurveyHumbatova, 2023SurveyMorovati, 2023SurveyChen, 2023SurveyHo, 2024SurveyTambon, 2024SurveyYu} manually categorize bug reports by root cause and symptom. They consistently identify training-phase algorithmic errors as primary, high-effort defects, mainly due to incorrect implementation of DL models and algorithms. Conversely, empirical research on DLCs is relatively sparse. Initial systematic analyses~\cite{2021SurveyShen, 2021SurveyDu} of mainstream DLCs~\cite{TVM2018, Glow2018, nGraph2018, plaidml, TC2018} identified tensor-type issues as a major root cause, which led to the development of TVMfuzz~\cite{2021SurveyDu}. More recently, Huang et al.~\cite{2025SurveyHuang} analyzed false-positive reports in TVM and OpenVINO~\cite{OpenVINO}.
In contrast to these labor-intensive, broad-scope surveys, our study is more focused and can achieve greater analytical depth and scalability.




\textbf{DL system testing}.
DL frameworks and DLCs are two main types of objects under test. 
Existing testing approaches for DL frameworks predominantly leverage fuzzing, ranging from API- and operator-level tests~\cite{2022DeepREL, 2022DocTer, 2022FreeFuzz, 2023ACETest}, which target API behaviors and individual operator implementations, to model-level tests~\cite{2019CRADLE, 2020Lemon, 2022Muffin, 2023NNSmith, 2023COMET, 2026SORT}, which synthesize diverse computation graphs to expose inconsistencies across frameworks. 
DLC testing evolves with the domain knowledge used. Early black-box strategies~\cite{2020TVMFuzz, 2022Tzer} mutate compiler IRs without internal insights. Subsequent approaches incorporate structural knowledge of computation graphs~\cite{2023GenCoG, 2024PolyJuice} or operator specifications~\cite{2025Opera} to guide test generation. More recently, white-box methods such as WhiteFox~\cite{2023Whitefox} and OATest~\cite{2025OATest} utilize LLMs to extract optimization semantics from source code for optimization-aware testing.
While progress has been made, current test generation relies on historical data. Our work, however, leverages in-depth analysis of the DLC mechanism to systematically characterize vulnerable compiler components and optimization stages, enabling LLMs to perform targeted testing.

\textbf{LLM-aided empirical study}.
Recent research uses LLMs to automate empirical studies~\cite{2024GPT4Empir, 2025LLMCluster, 2025AutoEmpirical}. Despite GPT-4’s identified limitations in domain-specific reasoning~\cite{2024GPT4Empir}, targeted pipelines like AutoEmpirical~\cite{2025AutoEmpirical} and LLMCluster~\cite{2025LLMCluster} successfully automate labor-intensive processes, including fault analysis and bug report clustering. Our work also benefits from these  advancements, i.e., significantly reducing manual effort in data preparation and analysis while maintaining high-quality insights.  


\textbf{Traditional compiler testing}.
Decades of research in conventional compiler testing, including random test case generation~\cite{2011CSmith, 2020YARPGen, 2024Fuzz4All}, optimization-focused testing~\cite{2014EMI, 2025Artemis}, and formal verification~\cite{2021Alive2, 2025Optimuzz}, cannot be directly adopted for DLC testing. The unique complexities of DLCs, particularly their tensor-centric computation and domain-specific optimizations, necessitate new testing paradigms beyond traditional approaches.

\section{Conclusion}
\label{sec:Conclusion}
This paper presents the first systematic empirical study of bugs in DLC frontends. Leveraging a domain-knowledge-enhanced LLM, we analyze 123 \emph{fBug}s and derive a bug root cause taxonomy comprising 7 categories and 15 subcategories. Besides, our study provides actionable insights for DLC development and testing. Beyond that, we uncover 23 new \emph{fBug}s (15 confirmed) in later releases, demonstrating its efficacy in guiding DLC testing.


\section*{Data Availability}
The source code and data are publicly available via Zenodo at \url{https://doi.org/10.5281/zenodo.19229496}.

\bibliographystyle{ACM-Reference-Format}
\bibliography{reference}

\end{document}